\documentclass[12pt,preprint]{aastex}

\def\etal{{et\,al.}}

\def\mdot{$\dot M$}

\def\msun{M$_{\odot}$}
\def\degs{\ifmmode ^{\circ}\else$^{\circ}$\fi}
\newbox\grsign \setbox\grsign=\hbox{$>$}
\newdimen\grdimen \grdimen=\ht\grsign
\newbox\laxbox \newbox\gaxbox
\setbox\gaxbox=\hbox{\raise.5ex\hbox{$>$}\llap
     {\lower.5ex\hbox{$\sim$}}}\ht1=\grdimen\dp1=0pt
\setbox\laxbox=\hbox{\raise.5ex\hbox{$<$}\llap
     {\lower.5ex\hbox{$\sim$}}}\ht2=\grdimen\dp2=0pt
\def\gax{$\mathrel{\copy\gaxbox}$}
\def\lax{$\mathrel{\copy\laxbox}$}

\shortauthors{Greiner et al.}
\shorttitle{GRB 080129}

\begin{document}

\title{A strong optical flare before the rising afterglow of GRB 080129}

\author{J. Greiner\altaffilmark{1}, T. Kr\"uhler\altaffilmark{1}, 
S. McBreen\altaffilmark{1}, M. Ajello\altaffilmark{1,2},
D. Giannios\altaffilmark{3},
R. Schwarz\altaffilmark{4},
S. Savaglio\altaffilmark{1},
A. K\"{u}pc\"{u} Yolda\c{s}\altaffilmark{5},
C. Clemens\altaffilmark{1},  
A. Stefanescu\altaffilmark{1}, G. Sala\altaffilmark{1,6}
%
%
F. Bertoldi\altaffilmark{7},
%
G. Szokoly\altaffilmark{8},
%
S. Klose\altaffilmark{9}}

\altaffiltext{1}{Max-Planck-Institut f\"ur Extraterrestrische Physik, 
         Giessenbachstra\ss{}e 1, 85740 Garching, Germany;
         jcg, kruehler, smcbreen, savaglio, cclemens, astefan@mpe.mpg.de}
\altaffiltext{2}{Present address: Stanford Univ., Stanford, CA 94305, U.S.A.;
       majello@slac.stanford.edu}

\altaffiltext{3}{Max-Planck-Institut f\"ur Astropysik, 85740 Garching,
  Germany; giannios@mpa-garching.mpg.de}

\altaffiltext{4}{Astrophysical Institute Potsdam, 14482 Potsdam, An der 
   Sternwarte 16, Germany; schwarz@aip.de}

\altaffiltext{5}{Present address: ESO, 85740 Garching, Schwarzschild-Str. 2,
  Germany; ayoldas@eso.org}

\altaffiltext{6}{Present address: Dept. de Fisica i Enginyeria Nuclear,
  EUETIB, Univ. Politecnica de Catalunya, c/ Compte d'Urgell 187,
  08036 Barcelona, Spain; gloria.sala@upc.edu}

\altaffiltext{7}{Astron. Inst. Univ. Bonn, Auf dem H\"ugel 71, 53121 Bonn, 
  Germany; bertoldi@astro.uni-bonn.de}
\altaffiltext{8}{E\"otv\"os Univ., 1117 Budapest, Pazmany P. stny. 1/A, 
  Hungary; szgyula@elte.hu}
\altaffiltext{9}{Th\"uringer Landessternwarte, Sternwarte 5, 07778 Tautenburg, Germany; klose@tls-tautenburg.de}


\begin{abstract}

We report on GROND observations of a 40 sec duration (rest-frame)
optical flare from GRB 080129 at redshift 4.349.
The rise-
and decay time follow a power law with indices +12 and -8, respectively,
inconsistent with a reverse shock
 and a factor 10$^5$ faster than variability caused by ISM interaction.
While optical flares have been seen in the past (e.g. GRB 990123, 041219B, 
060111B and 080319B),
for the first time, our observations not only resolve the optical flare
into sub-components, but also provide a spectral energy distribution 
from the optical to the near-infrared once every minute.
The delay of the flare relative to the GRB,
its spectral energy distribution as well as the ratio of pulse widths
 suggest it to arise from residual collisions in GRB outflows
\cite{liw08}.
If this interpretation is correct and can be supported by more detailed
modelling or observation in further GRBs,
the delay measurement provides an independent, 
determination of the Lorentz factor $\Gamma$ of the outflow.

\end{abstract}

\keywords{gamma-rays: bursts -- radiation mechanisms: non-thermal }


\section{Introduction}

Long-duration gamma-ray bursts (GRBs) emit their bulk luminosity over a
time period of 2-50 sec in the 100-1000 keV range (e.g. \cite{kaneko06}). 
Their afterglows are
generally assumed to arise from the interaction of the blast
wave with the surrounding interstellar material (ISM), where a strong
relativistic shock is driven (so-called external shock).
This happens about 10$^2$-10$^4$ sec after the burst, at distances
of the order of 3$\times$10$^{16}$ cm \cite{mep97}. 
The shocked gas is the source of a long-lived, slowly decaying
afterglow emission. 

Some afterglows have shown substantial optical variability, both 
at early times as well as at late times. The early ones can be distinguished 
into a component which tracks the prompt gamma-rays
(GRB 041219A \cite{vest05, bbs05}, GRB 050820A \cite{vest06}, 
 GRB 080319B \cite{rks08})
and an afterglow component which starts during or shortly after the
prompt phase 
(GRB 990123 \cite{akerlof99}, GRB 030418 \cite{ryk04}, 
GRB 060111B \cite{kgs06}).
The former component has been attributed to internal shocks,
while the latter component was interpreted as reverse shock emission,
e.g. \cite{sap99, mer99}.
At late times, some GRB afterglows (021004, 030329)
showed bumps on top of the canonical fading, with timescales of
10$^4$-10$^5$ sec. Originally, these bumps have  been interpreted as
the interaction of the fireball with moderate density enhancements in the 
ambient medium, with  a density contrast of order 10 \cite{lrc02},
and later by additional energy injection episodes \cite{bgj04}.

The optical variability due to the interaction with the ISM
is expected to be not faster than 10$^6$ sec, because the 
blast wave, once it has swept up enough interstellar material to produce the 
canonical afterglow emission, is thought to be only mildly relativistic.
This is different with optical emission possibly related to the
forward or reverse shock: here the emission is relativistic, 
and the timescales in the observer frame are shortened by 
$\Gamma^{-2}$, with $\Gamma$ being the bulk Lorentz factor
which typically is assumed to be 300--500.
The reverse shock is predicted to happen with little delay with respect to
the gamma-ray emission unless the Lorentz factor is very small, and 
the corresponding optical emission has a decay-time power law index of -2
for a constant density environment, or up to -2.8 for
a wind density profile \cite{kob00}.

Swift/BAT triggered on GRB 080129 (trigger 301981) at 06:06:46 UT
\cite{Immler08} which had an observed duration T$_{90}$=48 sec.
BAT measured a fluence (over T$_{90}$, the time
during which 90\% of the fluence is emitted) of
8.9$\times$10$^{-7}$ erg/cm$^2$ in the 15-150 keV
band\footnote{http://gcn.gsfc.nasa.gov/gcn/notices\_s/301981}.
The spectral slope is 1.3
with no spectral turn-over up to 150 keV. If we assume the expected
spectral turn-over according to a canonical GRB spectrum
to be at E$_{peak}$ = 300 (500) keV, the total isotropic gamma-ray 
energy equivalent is E$_{\gamma(iso)}$ = 6.5(7.7) $\times 10^{52}$ erg
(15--1000 keV).
At 320 sec after the trigger, Swift slewed to a different location on the sky,
placing the line-of-sight towards the
GRB nearly in the BAT detector plane, therefore being blind
to any late emission.
Pointed observations of the GRB with the X-ray telescope (XRT) and the
UV-optical telescope (UVOT) started only at 07:00:08 UT,
3.2 ksec after the GRB trigger. A clearly fading X-ray source
was discovered, but no emission seen with UVOT \cite{hol08}.

We started optical/near-infrared (NIR) imaging with GROND
immediately after the trigger, and
had independently identified the optical/NIR
afterglow \cite{klg08} though we reported it after 
Bloom (2008).
Here we report the full results.

\section{Observations and Results}

\subsection{Optical/NIR photometry}

GROND, a simultaneous 7-channel imager \cite{gbc08}
mounted at  the 2.2\,m MPI/ESO telescope at La Silla (Chile),
started observing the field at 06:10:18 UT, about 4 min after the GRB.
Our imaging sequence began with 46 sec integrations in the $g'r'i'z'$
channels, spaced at about 50 sec due to detector read-out and
preset to a new telescope dither position. After about 10 min,
the exposure time was increased to 137 sec, and after another 
28 min to 408 sec. Since the afterglow brightness was rising,
the exposure time was reduced back to 137 sec at 07:21 for the
rest of the night.
In parallel, the three near-infrared channels $JHK$ were
operated with 10 sec integrations, separated by 5 sec due to
read-out, data-transfer and $K$-band mirror movement.

\begin{figure}
\vspace{-1.cm}
\includegraphics[width=0.8\columnwidth]{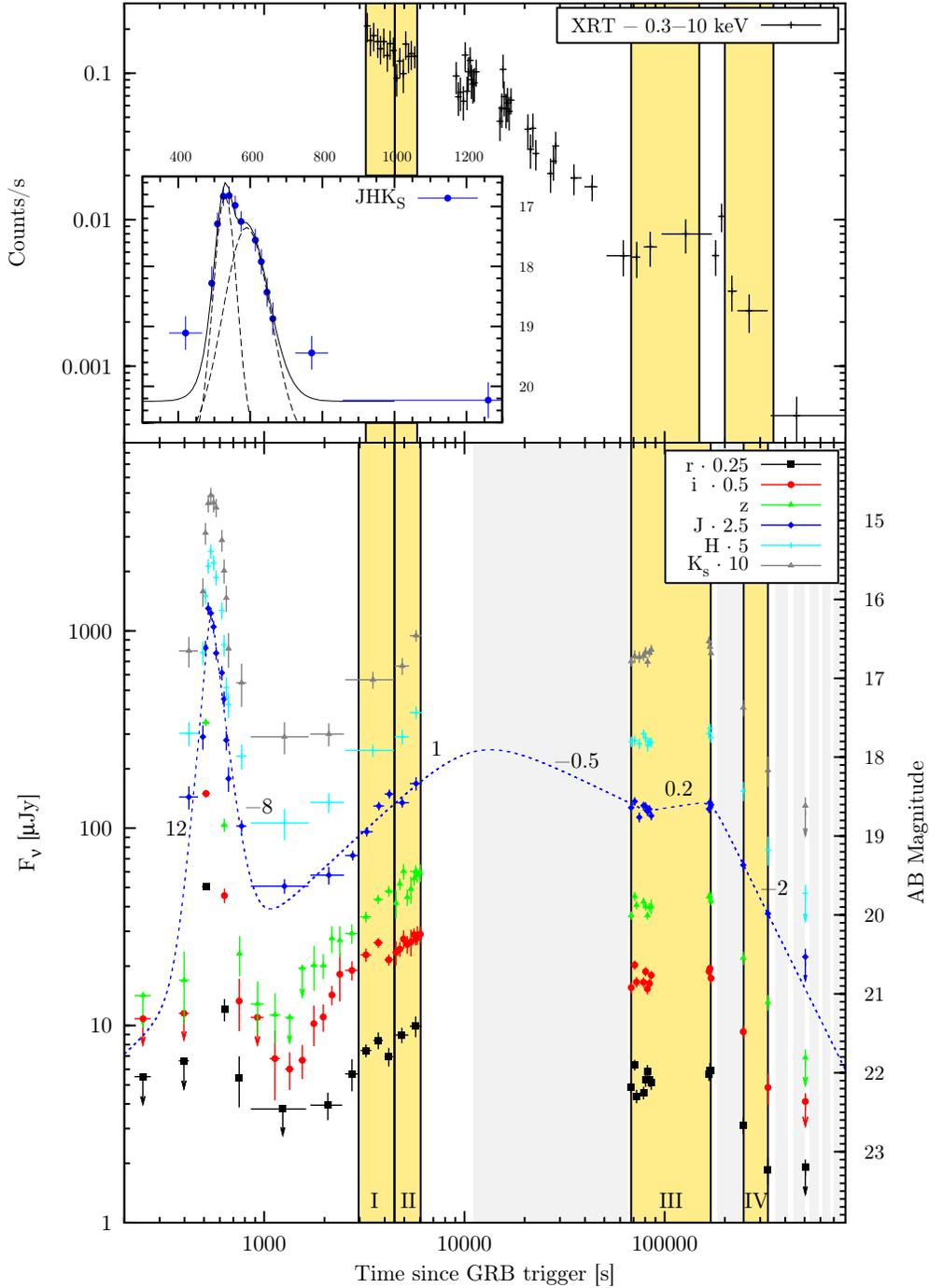}
 \vspace{-0.7cm}
   \caption[ospec]{\small Optical light curve of the afterglow of
     GRB 080129 obtained with the
     7-channel imager GROND at the 2.2m
    telescope on La Silla / Chile (bottom) and the X-ray light curve
    as measured with the XRT onboard Swift (top).
     NIR exposures  have been co-added until at least S/N=5$\sigma$
    was reached. During the optical/NIR flare at $\sim$500 sec, the individual
    10 sec integrations are shown. The inset in the top panel shows
    the flare in the three NIR channels co-added, and modelled by
    the sum (full line) of two (dashed lines) Gaussians with FWHM of
    77 and 157 sec, respectively.
    The $J$-band data have been fit by the sum of several power law
    segments shown as dashed line.  Numbers at this line indicate
    the temporal power law indices $\alpha$.
    The yellow-shaded areas are the time intervals of the SEDs
    as detailed in Tabs. \ref{SEDtab}, \ref{SEDtab2} and \ref{GRONDSEDtab}.
    \label{lc}}
\end{figure}

The first images
immediately revealed a strongly flaring source.
The light curve of the afterglow (Fig. \ref{lc}) shows this unique pattern
in more detail: there is a $\approx$3 mag amplitude flare of
80 sec (full-width at half maximum; FWHM) duration,
peaking at $\approx$540 sec post-burst.

Thereafter, the afterglow brightness is continuously rising until
6000 sec after the GRB. At the beginning of the next night, at
65 ksec after the GRB, the afterglow intensity is still at the same level,
despite declining by a factor of 25 at X-rays. In contrast, in the
65 - 500 ksec interval the emission in the optical/NIR and X-rays is
correlated, with a slow rise (t$^{0.15}$) over another day, and a
subsequent rapid decay (t$^{-2.0}$).

\subsection{Optical spectroscopy}

\begin{figure}[hb]
\includegraphics[angle=270, width=1.0\columnwidth]{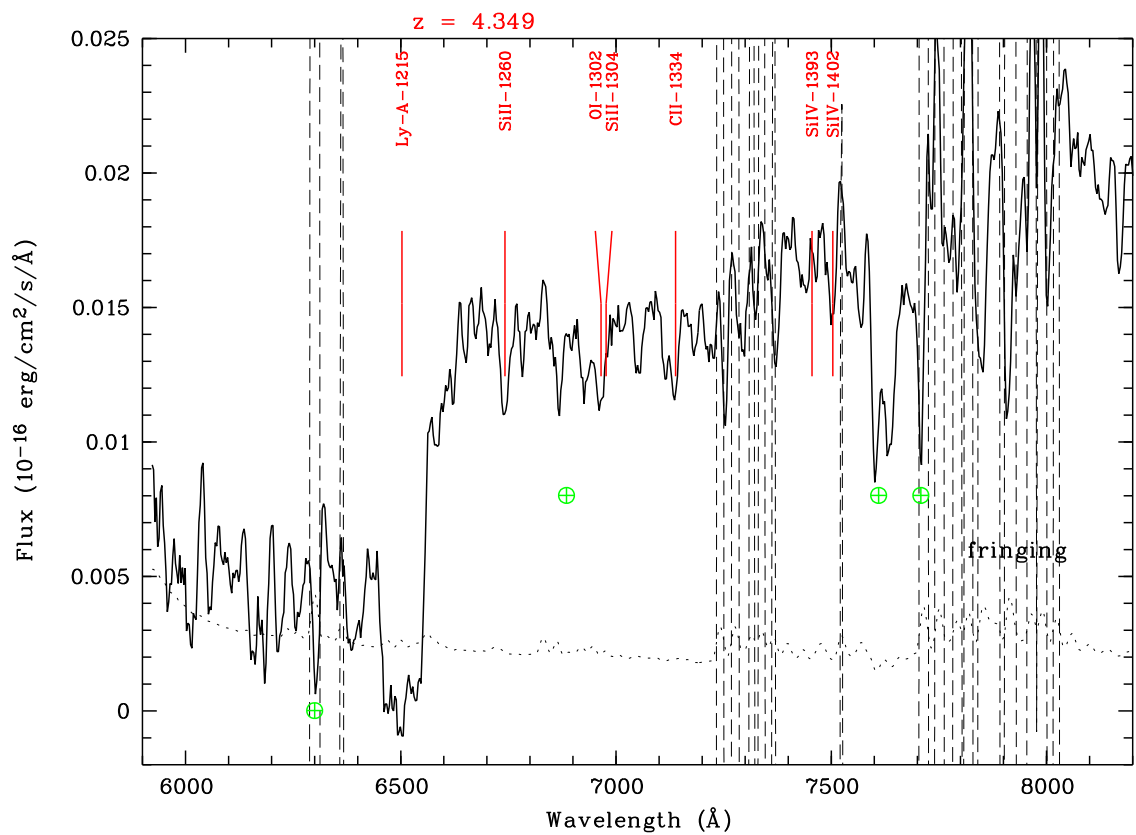}
    \caption[ospec]{Optical spectrum of the afterglow of GRB 080129
     obtained with FORS1/VLT
     on Jan 30, 2008.
    The Ly$\alpha$ line is clearly visible at 6500 \AA, and places
    GRB 080129 at a redshift $z$=4.349 (or larger). Some expected metal lines
    (in the rest frame) are indicated.
    The dotted line is the noise spectrum, and the vertical dashed lines
    mark regions of strong sky lines.
         }
    \label{redsh}
\end{figure}

We obtained an optical spectrum of the afterglow of GRB 080129
in the 500-800 nm region with FORS1/VLT
(Fig. \ref{redsh})  on Jan 30, 2008, 06:16 (mid-time)
consisting of 4 exposures of 1800 sec each.
The strong fringing of the blue-sensitive detector long-wards of 7500 \AA\
and the strong foreground extinction of $A_{\rm V}$=3.4 mag
result in a limited range of the spectrum being useful for analysis;
but luckily  Ly$\alpha$ and some metal absorption lines like
 SiII (1260 \AA) and SiIV (1402 \AA) happen to fall in this usable
range, so that
 a redshift of $z$=4.349$\pm$0.002 (luminosity distance of 40 Gpc
in concordance cosmology) could be derived.

\subsection{NIR high time-resolution photometry}

Observations with VLT/ISAAC (ESO Paranal, Chile) and 
NTT/SOFI (ESO La Silla, Chile) 
were triggered to monitor GRB 080129 in the NIR with high-time resolution
photometry. ISAAC was used in FastPhot mode in $J$ band on
2008 Jan 30, between UT 00:34 and 02:35, with 14.3 ms exposures, while
SOFI was used thereafter from UT 03:30 to 05:04 with
40.1 ms exposures (also FastPhotJitt mode in $J$ band).
After bias and flatfield correction and background subtraction
the frames were stacked to achieve longer total integration
times (4000 frames combined give 57.2 s integration time in ISAAC, 
1000 frames combined give 40 s integration time in SOFI). 
The light curves do not show any flaring activity above 0.3 mag amplitude.

\subsection{Sub-millimeter observations}

  For the photometric observations at 1.2 mm (250 GHz)  we used
the 117 channel Max-Planck Bolometer array MAMBO-2 \cite{kre98}
 at the IRAM 30 m telescope on Pico Veleta, Spain.
MAMBO-2 has a half-power spectral bandwidth from 210 to 290 GHz,
with an effective bandwidth center for flat spectra of
249$\pm$1 GHz (1.20 mm, 2$\pm$2mm percipitable water vapor). The
effective beam FWHM is 10.7 arcsec, and the undersampled field
of view is 4 arcmin.
Atmospheric conditions were generally good
during the observations, with typical line of sight opacities
between 0.2 and 0.3 and low sky noise.
The on sky integration
times varied between 1200 and 5000 sec on the five epochs.
Observations were performed using the standard on-off technique,
with the sub-reflector switching every 0.25 seconds between two
sky positions (on and off source) separated by 32 arcsec.
The telescope pointing was frequently checked on a nearby
quasar and was found to be stable
within 2 arcsec. 
The data were analyzed using the MOPSIC software package.
Correlated noise was subtracted from
each channel using the weighted average signals from the
surrounding channels.
Absolute flux calibration was done through observations of
planets, resulting in a flux calibration uncertainty of about
20\%. The third--fifth epochs on Feb 3, 6 and 10 yielded only
upper limits of $<$0.5 mJy (3$\sigma$) (see Tab. \ref{submm}).

\section{Discussion}

\subsection{The rising afterglow}

The rising light curve between 1000-6000 sec after the
GRB is likely the emerging afterglow. The rather steep power law photon
index of $\alpha$ = -1.35$\pm$0.15 and the flux rise (F $\sim$ t$^\beta$)
with $\beta \sim 1$ indicate
that the characteristic synchrotron frequency has already crossed the
optical band  at t=1000 sec. Our interpretation for the
rising part is that the ejecta have not entered the deceleration phase
at t=6000 sec. In this case one can use the peak time of the light curve at
t\gax 6000 sec, to estimate the fireball Lorentz factor at the time
of the deceleration which is expected to be half of the initial 
Lorentz factor $\Gamma_0$ \cite{sap99b, pak00, mol07}. 
Using the formulation of \cite{mol07}, we obtain for the ISM case
$\Gamma_0 \approx 130 \left
(\frac{E_{53}}{\eta_{0.2}n}\right)^{1/8}$,
where $\eta = 0.2 \eta_{0.2}$ is the radiative efficiency \cite{bfk03}.
Ignoring the weak dependence on $\eta$ and the external density $n$,
and using our above derived $E_{53}$=0.7,
we get $\Gamma_0 \approx 120$ (with allowed values down to 85 if the peak
emission was at 15.000 sec instead of 6.000 sec).

\subsection{The late decay light curve}

At very late times, starting at 180 ksec after the GRB, the X-ray and
optical/NIR emission vary achromatically. Again, this is in contrast
to the behaviour in most Swift GRBs \cite{panai08}, but the steepening of the
decay to $\alpha \sim -2$ and the spectrum by $\delta\alpha \sim 0.5$
(Tab. \ref{SEDtab}) is consistent with a jet break. 
The jet angle $\Theta$ was calculated following Sari et al. (1999) for
the ISM model and Bloom et al. (2003) for the wind model, where in the former
case the redshift factor was added:
\begin{eqnarray}
\Theta_{\rm ISM}  &=& \frac{1}{6}\ \left( \frac{t_b}{1+z}\right)^{3/8}\,
                      \left( \frac{n\,\eta_{0.2}}{E_{52}} \right)^{1/8}\\
\Theta_{\rm wind} &=& 0.169 \left( 2\, \frac{t_b}{1+z} \right)^{1/4}\,
                       \left( \frac{\eta_{0.2} A_\star}{E_{52}}\right)^{1/4}\,,
\end{eqnarray}
Using E$_{\rm iso}$ = 6.5(7.7)$\times 10^{52}$ erg/s (see introduction),
our derived redshift,
a circumburst density n = 1 cm$^{-3}$, and a break time of 
$t_{\rm b}$ = 180000 sec = 2.08 days, as well as the canonical
values $A_\star$ = 1 and $\eta_{0.2} = 1$, 
we derive a jet
opening angle of 4\fdg35(4\fdg26) for ISM and  3\fdg82(3\fdg66) for
a wind medium (where the density follows $A r^{-2}$, 
with $A$ = \mdot/4$\pi v = 5\times 10^{11} A_\star $ g cm$^{-1}$
derived for the reference values \mdot = 1$\times 10^{-5}$ \msun yr$^{-1}$
and $v = 1000$ km s$^{-1}$). 
The beaming factor is $b\approx\Theta^2/2$.
The corresponding jet angle-corrected energy is
1.88(2.13)$\times$10$^{50}$ erg/s for ISM, and
1.44(1.57)$\times$10$^{50}$ erg/s for wind medium.

\subsection{The plateau}

This GRB is remarkable for a second reason: it showed a prolonged
plateau phase in its afterglow emission, most pronounced in the X-ray band.
 Flat, or shallow-decay
parts of the light curve are now commonly detected in the Swift era
\cite{Liang07}, and occur between 100 sec until 10$^3$-10$^5$ sec after
the burst. In GRB 080129, we observe the plateau to last from
9000 - 56000 sec in the rest frame (50000 - 300.000 sec observers frame),
so starting substantially later, but with a duration (in the rest frame)
which is not extraordinary. However, the stunning fact is that this same
plateau is also seen in the optical/NIR data of GROND. Using also
the MAMBO detection at 1.2 mm, the overall spectrum during the plateau
cannot be fit by a single power law, but requires a second component.
Adopting a broken power law, at least one break is required, with
the break energy between the
optical (400 nm) and X-rays (0.5 keV).
The best-fit power law indices are 1.57$\pm$0.06 for the MAMBO-GROND spectrum,
and 2.36$^{+1.01}_{-0.58}$ for the high-frequency part of the spectrum.
Integrating this spectrum over the
duration of the plateau phase (69 hrs) results in a total emitted energy,
of 3.4$\times 10^{52}$ erg, about 50\% of the total energy emitted during
T$_{90}$ in the 15--150 keV band.

\subsection{The flare}

\subsubsection{Non-favored explanations}

The optical flare is more difficult to explain due to primarily
two facts: it is not correlated to the gamma-ray emission, but delayed
by 12$\times$T$_{90}$,
and it occurs well before the peak of the optical afterglow.
One possibility is to assume that it is
the prompt emission of the GRB while BAT triggered on the precursor.
The typical ratio of at least 30 for the gamma-ray fluence of proper burst to
precursor \cite{laz05} implies 
E$_{\gamma(iso)}$ = 2.0(2.3) $\times$ 10$^{54}$ erg,
similar to the brightest previously known burst GRB 990123 \cite{akerlof99}, 
therefore making the precursor hypothesis unlikely.

Another possibility to explain the optical flare is 
as the reverse shock emission. 
In a constant density environment, a reverse shock \cite{kob00} is expected
to rise rapidly ($\beta_{rise} = 3p - 3/2$, where p is the powerlaw index
of the electron distribution), and decline, in the thin shell case,
with $\beta_{decline} = - (27p + 7)/35$. With the canonical range of 
$p = 2.2-2.5$, this implies $\beta_{rise} = 5.1 - 6.0$,
and $\beta_{decline} = -1.9 - -2.1$, in contrast to our observed
values of $\beta_{rise} = 12.1\pm1.5$ and $\beta_{decline} = 8.3\pm1.8$
While this is true only for the simplest model, and the actual
rise and decline values depend on the density profile and the $p$ values
of the electron distribution, we are not aware of any reverse shock model
that gives so steep flux density variations.
Note also that a wind profile, while helping in steepening the decline time,
would not give a rising forward shock optical emission as we observe.

Yet another option is to interpret the flare as the simultaneous optical
emission from an unobserved (because Swift/XRT did not point to 
the GRB at that time) X-ray flare. X-ray flares are commonly seen
in GRB afterglows, at times typically 1000-10000 sec (rest frame) after the GRB
\cite{cmr07}.
In our case, the early occurrence would be on the short side of this
distribution, still consistent with this distribution.
The presently generally accepted explanation for the X-ray flares is
that they are due to late-time   internal shocks \cite{koc07},
in particular either with a low $\Gamma$-difference
(so they collide late), or ejected with a large time difference
(late-time activity of the engine). For both cases, one expects
that the rise time is (much) shorter than the decay time:
     The rise time is basically the time it takes for the reverse
     shock of that collision to travel through the thickness of the
     shell. The decay time is due to the curvature effect, becoming
     important whenever the radius of the shell exceeds the shell thickness.
     Thus, if we require that the decay time is not larger than
     the rise time (as we observe), then the shell radius must be
     of the order of the shell thickness - and this is valid only
     very early after the GRB, thus incompatible with our late-time
     occurrence.
Also, simultaneous Swift/UVOT observations of the many X-ray flares
have not revealed such flaring activity in the UV/optical domain.
Thus,
we consider it unlikely that the optical flare in GRB 080129 is the optical
counterpart of an unseen X-ray flare.

Invoking a late internal shock between
shells which have not produced gamma-ray emission, and collide
at large radii, is another option. While this
scenario has been already proposed to explain the early optical emission
of GRBs 990123, 041219 and 060111B \cite{wei07}, it 
requires that the late ejections have, for some reason, 
very high $\Gamma$ of order 800--1000, without producing gamma-ray emission.
This $\Gamma$ value is well above the
measured $\Gamma$ \lax 120 of the main burst.

Finally, our light curve has, at first glance, some resemblance to that 
of GRB 041219A, a long-duration ($T_{90}=520$ sec) burst for which
PAIRITEL obtained infrared photometry starting before the end
of the burst \cite{bbs05}.
In that case, the first flare, occuring before the end of the
burst emission, was associated to the internal shock that produced the GRB;
however, the note added in proof implies that a re-analysis
of the data showed less evidence for the rising part. Thus, it remains
open whether this emission was indeed a flare, or some slower-decaying
prompt emission.
The second flare at 3$\times T_{90}$ was associated with the reverse shock.
The rise and decline times of this second flare,
$\beta_{rise} = 6.1\pm2.9$ and $\beta_{decline} = -3.4\pm2.8$  are
fully consistent even with the simplest model of a reverse shock,
while our values for GRB080129 are not.
Thus, the similarity between the observed optical/infrared light curves
of GRB 041219 and 080129 ends with the global structure of multiple
peaks in the light curves, but does not provide clues to solve the
discrepancies in the case of GRB 080129.

\subsubsection{The likely cause of the flare}

The best match of the observed properties of the flare in GRB 080129
with theoretical predictions is with residual collisions in GRB
outflows \cite{liw08}. Internal collisions at small radii, which produce
the $\gamma$-ray emission, have been proposed to lead to residual collisions
at much larger radii where the optical depth to long-wavelength photons
is much lower. If the bulk Lorentz factor is large, the optical emission
is delayed by only fractions of a second with respect to the $\gamma$-rays,
and thus can explain the prompt optical emission which has been seen 
so far in a few GRBs like GRB 041219A, 050820A or 080319B.
In the case of GRB 080129, $\Gamma$ \lax 120, and the delay time
can be longer than the duration of the burst (in such case 
the electrons that radiate in the optically emitting region
do not cool because of up-scattering the GRB photons). Li \& Waxman (2008) 
showed
that the radius at which  the (observer-frame) NIR $\sim$10$^{14}$ Hz radiation
becomes optically thin is R$_{\rm NIR} \sim 7.3\times 10^{15}
L_{k,52}^{1/2}\Gamma_2^{1/2}$ cm, resulting in a delay
$\tau \sim R_{\rm NIR}/2\Gamma^2 \sim 12 L_{k,52}^{1/2}/\Gamma_2^{3/2}$ sec. 
Assuming that the kinetic luminosity of the flow 
L$_{\rm k} \sim 10 L_\gamma \sim 10^{53}$ erg/sec (in
fact the very long phase of optical emission between 1-3 days after the GRB
with a luminosity similar to that of the burst itself implies a
 large kinetic energy), the delay time can be $\tau \sim$ 100 sec if
$\Gamma \sim 50$. The predicted spectral slope above the self-absorption
frequency is $\nu F_\nu \sim \nu^{0.5}$, and $\nu^{7/6}$
below, consistent with our measured values of 0.57 and 1.2,
 respectively (rising part of the flare).
Also, the predicted ratio of
F$_\gamma$/F$_{opt} \sim 500$ compares well with the observed ratio of
1000.

\begin{figure}
\includegraphics[width=0.75\columnwidth,angle=270]{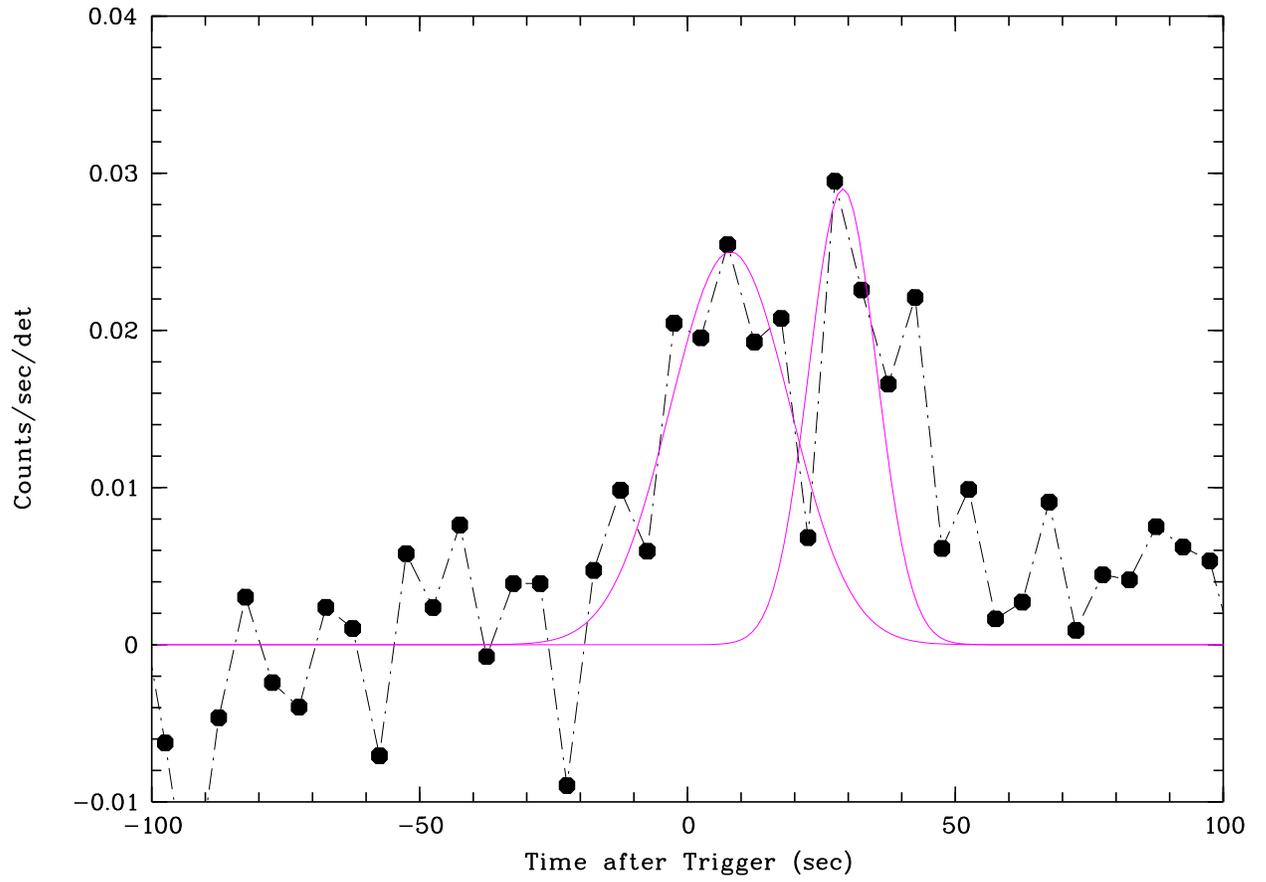}
  \caption[oima]{Swift/BAT light curve of GRB 080129, rebinned with
    S/N=5. Overplotted are the two peaks, modelled with two Gaussians
    of 11 sec and 6 sec FWHM, respectively.
    \label{bat}}
\end{figure}

Given this tantalizing coincidences, we analyzed in more detail the shape
of the optical flare light curve. It turns out that it can be well
described by the superposition of two Gaussian profiles (Fig. \ref{lc}; note
that log Gaussians or fast-rise-exponential-decay curves do not fit).
We speculate that these are the direct signatures of the residual collisions.
Looking at the $\gamma$-ray light curve from Swift/BAT (Fig. \ref{bat}), 
one can recognize
two pulses, the first with FWHM = 11 sec, the second with FWHM = 6 sec.
It is interesting to note that the ratio of the FWHM of these pulses
is two, identical to the corresponding ratio of the optical pulses.
Given that just the sequence of broad/narrow pulse has inverted, one
could speculate even further that the shell causing the narrow, second
peak in $\gamma$-rays had a slightly higher $\Gamma$ and took over the
shell causing the broader, first $\gamma$-ray pulse, thus leading
to the optical flare.


A caveat with this interpretation comes from the observed fast
variability of the flare. Residual collisions are expected to result in a
smooth optical lightcurve that varies on the delay timescale.
Alternatively the flare may be powered by dissipation of
Poynting flux in a localized ``hot spot'' in strongly magnetized ejecta
\cite{Lyut06, Giann06}.
In this picture the fast variability is the result of the small emitting
volume.
The observed fluence of the flare is comparable to the energy available
in the volume of the hot spot as constrained by the observed fractional
 duration of the flare
$\delta t_{\rm f}/t_{\rm f} \sim 0.15$ \cite{Giann06}. 
The energy contained in the ``hot spot'' is
$E_{\rm HS} \sim E_{\gamma,iso}(\delta t_f/t_f)^3 \sim 3\times 10^{50}$ erg 
(assuming
again that the total energy in the ejecta is $\sim$10 times larger than the
$E_{\gamma,iso}$)
In this scenario of a ``hot spot'',
the radiation would also be strongly polarized - a prediction
which can help to distinguish the above two models by future
observations of similar phenomena.

\section{Conclusions}


If more detailed theoretical investigation of the properties of residual 
collisions and the comparison of their predictions with our data will support 
our interpretation of the observed flare to be correct, 
then the delay time between gamma-ray and optical
flare provides an independent way of determining
the Lorentz factor $\Gamma$.
 Moreover, further parameters of the blast wave
can be determined, which were not constrained by observations so far,
such as the distance of the residual
collisions, the ratio of radiation to magnetic field energy (via the
ratio of inverse Compton and synchrotron emission), and the ratio
of kinetic to gamma-ray energy.
This offers the hope to finally measure the energetics
of gamma-ray bursts beyond the rare cases of calorimetry with
radio observations.



\acknowledgements 

We are grateful to  Pierre Cox, the IRAM Director, for granting
DDT time at the 30m telescope,
as well as to C. Thum and  S. Leon (also IRAM) for 
getting the observations performed.
This work is partly
based on observations collected at the European Southern Obser\-vatory,
Chile under proposal ESO No. 280.D-5059.




\newpage



\begin{table}[ht]
\caption[sub-mm]{Sub-mm measurements of GRB 080129 with MAMBO.\label{submm}}
  \begin{tabular}{crcrcccc}
   \noalign{\smallskip}
  \hline
  \noalign{\smallskip}
  Date     &  LST & MJD     & Flux~~~   & Exp.time& Opacity  & Elev & Scan \\
           & (hr) &         & (mJy)     & (sec)   &          & (deg) &  \\
  \noalign{\smallskip}
  \hline
  \noalign{\smallskip}
 2008-01-30&  6.1 &54495.90 & 2.98$\pm$0.63&  2358  &0.20--0.21& 43-45 &1-2 \\
 2008-01-31&  7.9 &54496.98 & 1.27$\pm$0.47&  5097  &0.23--0.29& 43-29 &3-7 \\
 2008-02-03&  9.3 &54499.02 & 0.50$\pm$1.16&  1415  &0.27      & 35-33 &8-9 \\
 2008-02-06&  8.6 &54502.99 &-0.40$\pm$0.55&  3540  &0.20--0.24& 30-40 &10-12\\
 2008-02-10& 10.4 &54506.05 & 0.55$\pm$1.14&  1170  &0.29--0.33& 22-25 &13\\
  \noalign{\smallskip}
  \hline
  \end{tabular}
\end{table}

\clearpage

\begin{table}[ht]
\caption[SED]{Fit parameters of the combined Swift-XRT/GROND/MAMBO
  SEDs. The normalisation is in ph/cm$^2$/s/keV, and only a break between
  GROND and XRT data is fit, with the break energy
  fixed at 0.1 keV. Leaving the break energy free
  results in  best-fit values between 0.05-0.25 keV with large errors.
  We have no evidence that
  the break energy moved  in time, neither between the
  GROND and XRT bands, nor through the GROND band towards shorter frequencies.
\label{SEDtab}}
  \bigskip
  {\small
  \begin{tabular}{cccccc}
  \hline
  \noalign{\smallskip}
   Interval & Time (ksec post-GRB) &  ~Low-energy power  &
    ~High-energy power & Norm & $\chi^2_{\rm red}$/d.o.f \\
              & ~GROND/XRT/MAMBO~& law photon index &  law photon index & & \\
  \noalign{\smallskip}
  \hline
  \noalign{\smallskip}
  1 &2.96-4.46/3.22-4.48/--  & 1.41$\pm$0.13 &    2.12$\pm$0.33 &   6.45E-03
& 0.94/10     \\
  2 &4.48-6.04/4.48-5.80/--  & 1.39$\pm$0.11 &    2.33$\pm$0.38 & 5.93E-03 & 
0.97/7 \\
  3 &68-170/68-149/141-143   & 1.57$\pm$0.06 &    2.36$^{+1.01}_{-0.58}$ &  1.41E-03    & 0.81/3 \\
  4 &248-328/250-350/232-237 & 1.60$\pm$0.09 &    1.92$^{+1.50}_{-0.10}$&  
4.93E-04 & 0.41/4  \\
  \noalign{\smallskip}
  \hline
  \end{tabular}}
\end{table}

\clearpage

\begin{table}[ht]
\caption[SED]{Fit parameters of the combined Swift-XRT/GROND/MAMBO
  SEDs when enforcing two breaks, one between MAMBO and GROND, and
  the other one between GROND and Swift/XRT.
  The normalisation is in ph/cm$^2$/s/keV.
  The break energies have been fixed at 5E-5 keV (1.2 mm) and 0.5 keV,
  respectively.
\label{SEDtab2}}
  \bigskip
  {\small
  \begin{tabular}{cccccc}
  \hline
  \noalign{\smallskip}
SED  &   Gamma1  & Gamma2& Gamma3 & Norm  & $\chi^2_{\rm red}$/d.o.f \\
  \noalign{\smallskip}
\hline
  \noalign{\smallskip}
SED III & 1.47$\pm{0.14}$ & 1.69$\pm{0.13}$& 2.31$\pm{0.23}$ & 5.97e-3 & 0.38/4\\
SED IV & 1.53$\pm{0.24}$ & 1.67$\pm{0.21}$& 2.10$\pm{0.38}$ & 1.08e-3 & 0.43/4\\
  \noalign{\smallskip}
  \hline
  \end{tabular}}
\end{table}

\clearpage

\begin{table}[ht]
\caption[SED]{Fit parameters of the GROND SED data alone during the
4 intervals as shown in Fig. \ref{SED}.
  For the flare SEDs a break is required, and has been fixed at 1500 nm.
 $\beta_1$ and $\beta_2$ are the low- and high-energy pwer law photon indices,
 respectively.
 \label{GRONDSEDtab}}
 \bigskip
  {\small
  \begin{tabular}{ccccc}
  \hline
  \noalign{\smallskip}
    Time  & $\beta_1$  & Break &  $\beta_2$  & Norm \\
                    &  & nm &   & $\mu$Jy\\
  \noalign{\smallskip}
  \hline
  \noalign{\smallskip}
   1.  flare peak       & 0.57$\pm$0.27 & 1500 & 1.20$\pm$0.10 & 363$\pm$17 \\
   2.  flare decay      & 0.99$\pm$0.26 & 1500 & 1.87$\pm$0.09 & 164$\pm$7\\
   3.  rising AG & 1.27$\pm$0.04 &  --  & -- & 10.9$\pm$ 2.4 \\
   4.  peak AG   & 1.52$\pm$0.04 &  --  & -- & 4.2$\pm$0.9 \\
  \noalign{\smallskip}
  \hline
  \end{tabular}}
\end{table}

\newpage
 

\begin{figure}
\includegraphics[angle=270, width=1.\columnwidth]{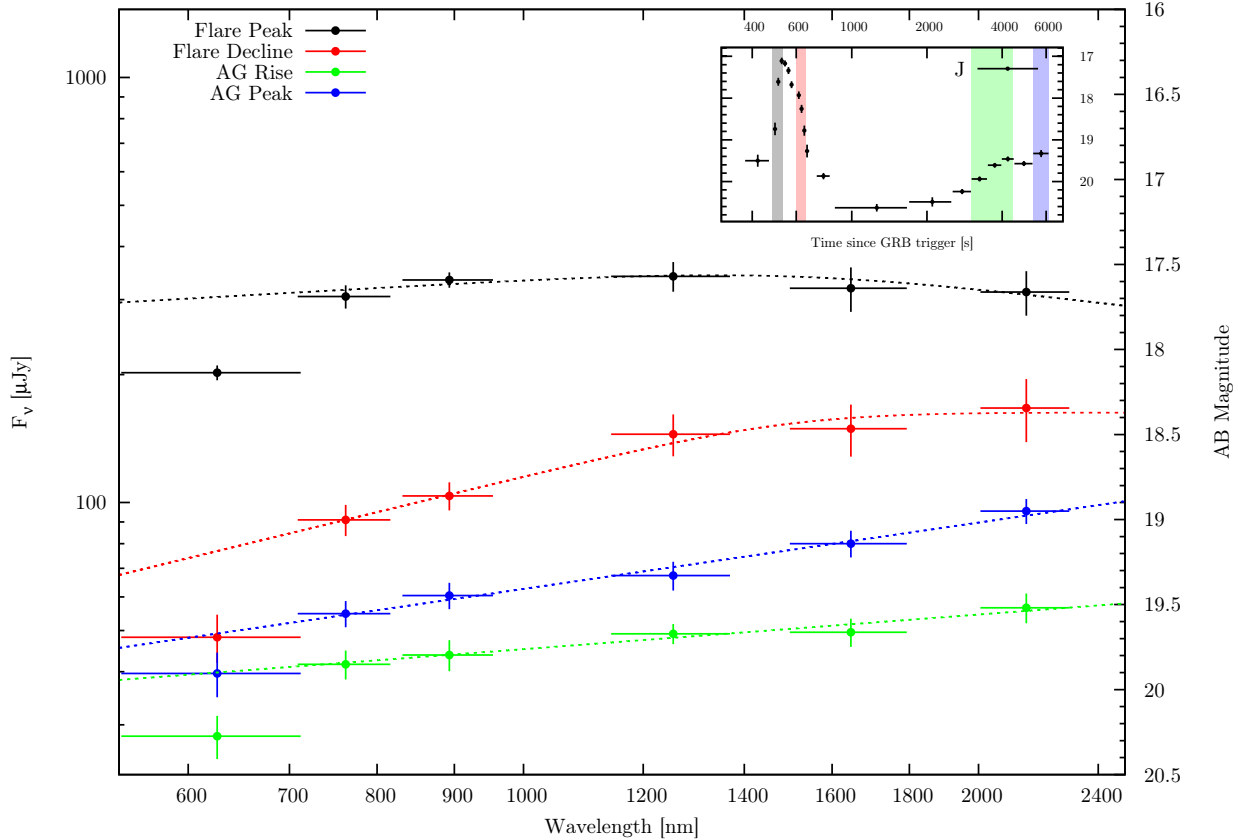}
    \caption[SEDs]{Spectral energy distribution (SED) at four different
   times (see inset; from top to bottom):
   (i) the peak of the flare
   (ii) the decay part of the flare
   (iii) the maximum of the afterglow emission, and
   (iv) the rising part of the afterglow emission
   as measured by GROND.
  The Ly-$\alpha$ line affects the $r'$-band, and has not been included
  in the fit.
  The burst location is at galactic latitude -1.42 deg, thus the foreground
  galactic hydrogen column is N$_{\rm H}$ = (6.7-7.5)*10$^{21}$ cm$^{-2}$
 \cite{DL90}, corresponding to a  Galactic visual extinction of
  A$_V$ = 3.5-4.1 mag.
  Our best-fit extinction is A$_V$ = 3.4 mag, and all measured magnitudes
  have been corrected for this extinction.
  The power law photon indices are given in Tab. \ref{GRONDSEDtab}.
  The break in the flare spectrum
   violates one of the assumptions made for deriving the $\alpha$ = 2 + $\beta$
   relation for the curvature effect; thus, the curvature effect can not be
   tested for the observed optical flare.
         }
    \label{SED}
\end{figure}


\begin{figure}
\includegraphics[width=0.65\columnwidth, angle=270]{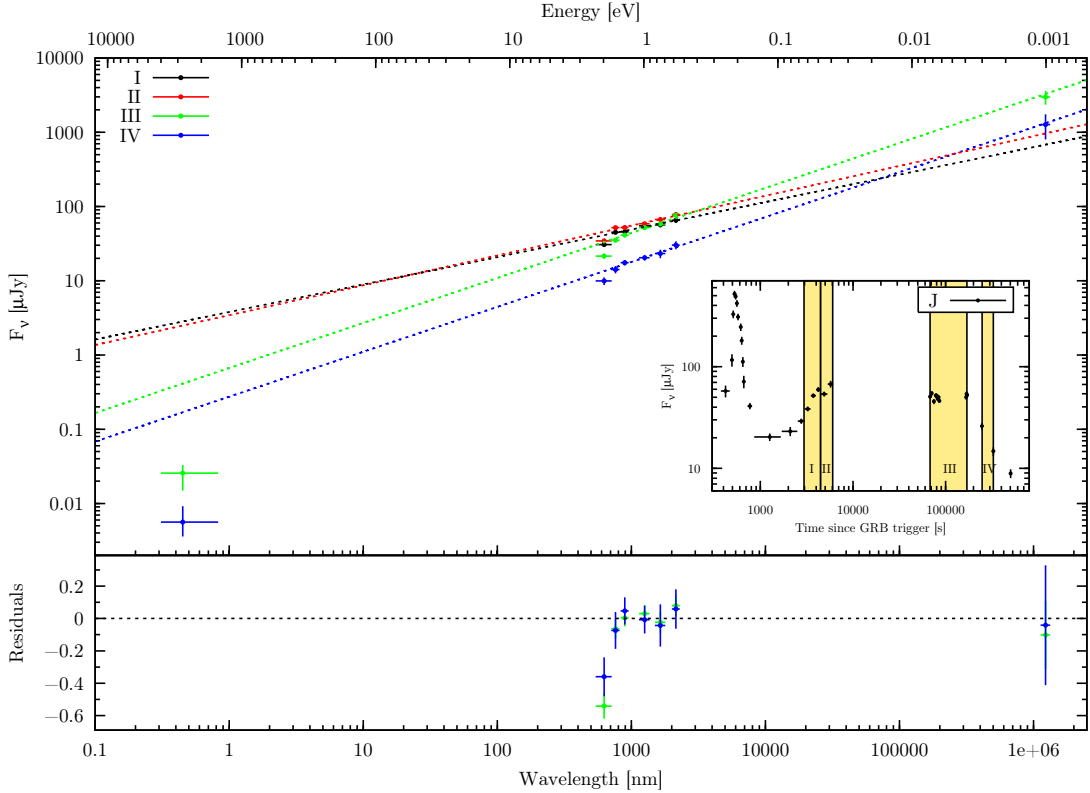}
    \caption[SED]{\small Broad-band spectrum of GRB 080129 at different
    epochs (see inset and legend),
    combining GROND data (center) with Swift/XRT (left) and
    MAMBO (top right). The best-fit photon indices are 2.66$\pm$0.12 at
   wavelengths shorter than 400 nm, and 1.62$\pm$0.03 above. The best-fit
   extinction of the optical/NIR fluxes is A$_V$=3.4$\pm$0.1 mag, and the
   neutral hydrogen absorption N$_{\rm H} =6 \times 10^{21}$ cm$^{-2}$ which
   are nicely consistent with the canonical galactic conversion.
   The X-ray data are a factor $\sim$10 below the power law connecting
   GROND and MAMBO, and thus a break in the spectrum is required.
         }
    \label{broadSED}
\end{figure}


\end{document}